\documentstyle[12pt,subeqn,subeqnarray,epsf,epsfig,fullpage]{article}

\def\be{\begin{equation}}
\def\ee{\end{equation}}
\def\beq{\begin{eqnarray}}
\def\eeq{\end{eqnarray}}
\def\lsim{\:\raisebox{-0.5ex}{$\stackrel{\textstyle<}{\sim}$}\:}
\def\gsim{\:\raisebox{-0.5ex}{$\stackrel{\textstyle>}{\sim}$}\:} 
\def\GeV{{\rm ~GeV}}

\begin{document}
\hfill TIFR/TH/99-05
\bigskip
\begin{center}
{\Large{\bf SUSY \& SUGRA SIGNATURES AT HADRON
COLLIDERS}\footnote{Invited talk at the 13th Intl. Conf. on Hadron
Collider Physics, Mumbai, 14-20 January 1999.}} \\[1cm]
{\large D.P. Roy} \\[.5cm]
Tata Institute of Fundamental Research, \\ Homi Bhabha Road, 
Mumbai 400 005, India
\end{center}
\bigskip\bigskip

\noindent {\bf Abstract:} After a brief introduction to SUSY I discuss
the missing-$p_T$ signature for superparticles from $R$-parity
conservation and the multilepton signature, which follows from their
cascade decay.  The GUT and SUGRA constraints on the SUSY mass
parameters are discussed along with the resulting SUSY signals at
LHC.  Finally I consider the effect of relaxing the SUGRA constraint
on these signals.
\bigskip\bigskip

\noindent {\bf Why SUSY? (Hierarchy Problem):} Assuming the Higgs
mechanism  of electroweak symmetry breaking one is
faced with the hierarchy problem, i.e. how to peg down the Higgs
scalar in the desired 
mass range of $\sim 10^2 \GeV$.  This is because the scalar
masses are known to have quadratically divergent quantum corrections
from radiative loops involving e.g. scalars or fermions.  These would
push the output scalar mass to the cut-off scale of the SM, i.e. the
GUT scale $(10^{16} \GeV)$ or the Planck scale $(10^{19} \GeV)$.  The
desired mass range of $\sim 10^2 \GeV$ is clearly tiny compared to
these scales.  The underlying reason for the quadratic divergence is
that the scalar masses are not protected by any symmetry unlike the
fermion and the gauge boson masses, which are protected by chiral
symmetry and gauge symmetry.  Of course it was this very property of
the scalar mass that was exploited to give masses to the fermions and
gauge bosons in the first place.  Therefore it can not be simply
wished away.

The most attractive solution to this problem is provided by
supersymmetry (SUSY), a symmetry between fermions and bosons [1].  It
predicts the quarks and leptons to have scalar superpartners called
squarks and sleptons $(\tilde q, \tilde \ell)$, and the gauge bosons
to have fermionic superpartners called gauginos $(\tilde g,\tilde
\gamma, \tilde W, \tilde Z)$.  In the minimal supersymmetric
standard model (MSSM) one needs two Higgs doublets $H_{1,2}$,
with opposite hypercharge $Y = \pm 1$, to give masses to the up and
down type quarks. The ratio of their vevs is denoted by $\tan\beta$. The corresponding fermionic superpartners are
called Higgsinos $(\tilde H_{1,2})$.  The opposite hypercharge of
these two sets of fermions ensures anomaly cancellation.

SUSY ensures that the quadratically divergent quantum corrections from
scalar and fermion loops are cancelled by the loop contributions from the
corresponding super partners.  Thus the Higgs masses can be
kept in the desired range of $\sim 10^2 \GeV$.  However this implies
two important constraints on SUSY breaking. 

\begin{enumerate}
\item[{i)}] SUSY can be broken in masses but not in couplings (soft
breaking), so that the co-efficients of the cancelling contributions
remain equal and opposite.
\item[{ii)}] The size of SUSY breaking in masses is $\sim 10^2 \GeV$,
so that the size of the remainder remains within this range.  Thus the
superpartners of the SM particles are also expected to lie in the mass
range of $\sim 10^2 \GeV$, going upto $1000 \GeV$.
\end{enumerate}
\bigskip

\noindent {\bf $R$-Parity Conservation \& the Missing-$p_T$
Signature:}  I shall concentrate on the standard $R$-Parity conserving
SUSY model.  Let me start therefore with a brief discussion of
$R$-parity.  The presence 
of scalar quarks in SUSY can lead to baryon and lepton number
violating interactions of the type $ud \rightarrow \bar{\tilde s}$ and
$\bar {\tilde s} \rightarrow e^+ \bar u$, i.e.
\be
u d \ {\buildrel {\bar{\tilde s}} \over \longrightarrow} \ e^+ \bar u.
\ee
Moreover adding a spectator $u$ quark to both sides one sees that this
can lead to a catastrophic proton decay, i.e. 
\be
p (uud) \ {\buildrel {\bar{\tilde s}} \over \longrightarrow} \ e^+ \pi^0
(\bar u u).
\ee
Since the superparticle masses are assumed to be of the order $M_W$
for solving the hierarchy problem, this would imply a proton life time
similar to the typical weak decay time of $\sim 10^{-8} {\rm sec}$!
The best way to avoid this catastrophic proton decay is via $R$-parity
conservation, where
\be
R = (-1)^{3B+L+2S}
\ee
is defined to be $+1$ for the SM particles and $-1$ for their
superpartners, since they differ by $1/2$ unit of spin $S$.  It
automatically ensures $L$ and $B$ conservation by preventing single
emission (absorption) of superparticle.

Thus $R$-conservation implies that (i) superparticles are produced in
pair and (ii) the lightest superparticle (LSP) is stable.
Astrophysical evidences against such a stable particle carrying colour
or electric charge imply that the LSP is either sneutrino $\tilde \nu$
or photino $\tilde\gamma$ (or in general the lightest neutralino).
The latter alternative is favoured by the present SUSY models.  In
either case the LSP is expected to have only weak interaction with
ordinary matter like the neutrino, since e.g.
\be
\tilde\gamma q \ {\buildrel {\tilde q} \over \longrightarrow} \ q
\tilde \gamma \ \ {\rm and} \ \ \nu q \ {\buildrel W \over
\longrightarrow} \ e q'
\ee
have both electroweak couplings and $M_{\tilde q} \sim M_W$.  This
makes the LSP an ideal candidate for the cold dark matter.  It also
implies that the LSP would leave the normal detectors without a trace
like the neutrino.  The resulting imbalance in the visible momentum
constitutes the canonical missing transverse-momentum $(p\!\!\!/_T)$
signature for superparticle production at hadron colliders.  It is
also called the missing transverse-energy $(E\!\!\!/_T)$ as it is
often measured as a vector sum of the calorimetric energy deposits in
the transverse plane.

The main processes of superparticle production at hadron colliders are the QCD
processes of quark-antiquark and gluon-gluon fusion [2]
\be
q\bar q, gg \longrightarrow \tilde q \bar{\tilde q} (\tilde g \tilde
g).
\ee
The NLO corrections can increase these cross-sections by $15-20\%$
[3].  The simplest decay processes for the produced squarks and
gluinos are
\be
\tilde q \rightarrow q \tilde \gamma, \ \tilde g \rightarrow q\bar q
\tilde \gamma.
\ee
Convoluting these with the pair production cross-sections (5) gives
the simplest jets + $p\!\!\!/_T$ signature for squark/gluino
production, which were adequate for the early searches for relatively
light squarks and gluinos.  
\bigskip

\noindent {\bf Cascade Decay and the Multilepton Signature:} Over the
mass range of current 
interest $(\geq 100 \GeV)$ however the cascade decays of squark and gluino
into the LSP via the heavier chargino/neutralino states are expected
to dominate over the direct decays (6).  This is both good news and
bad news.  On the one hand the cascade decay degrades the missing-$p_T$
of the canonical jets $+ p\!\!\!/_T$ signature.  But on the other hand
it gives a new multilepton signature via the leptonic decays of these
chargino/neutralino states.  It may be noted here that one gets a mass
limit of
\be
M_{\tilde q,\tilde g} > 180 \GeV
\ee
from the Tevatron data using either of the two signatures [4].

The cascade decay is described in terms of the $SU(2) \times U(1)$
gauginos $\tilde W^{\pm,0}, \tilde B^0$ along with the Higgsinos
$\tilde H^\pm$, $\tilde H^0_1$ and $\tilde H^0_2$.  The $\tilde B$ and
$\tilde W$ masses are denoted by $M_1$ and $M_2$ respectively while
the Higgsino masses are functions of the supersymmetric Higgsino mass
parameter $\mu$ and $\tan\beta$.  The charged and the neutral gauginos
will mix with the corresponding Higgsinos to give the physical
chargino $\chi^\pm_{1,2}$ and neutralino $\chi^0_{1,2,3,4}$ states.
Their masses and compositions can be found by diagonalising the
corresponding mass matrices, i.e.
\[
M_C = \left(\matrix{M_2 & \sqrt{2} M_W \sin \beta \cr & \cr \sqrt{2}
M_W \cos \beta & \mu}\right),
\]
\bigskip
\be
M_N = \left(\matrix{M_1 & 0 & -M_Z \sin \theta_W \cos \beta & M_Z \sin
\theta_W \sin \beta \cr & & & \cr 0 & M_2 & M_Z \cos \theta_W \cos
\beta & -M_Z \cos \theta_W \sin \beta \cr & & & \cr -M_Z \sin \theta_W
\cos \beta & M_Z \cos \theta_W \cos \beta & 0 & -\mu \cr & & & \cr M_Z
\sin \theta_W \sin \beta & -M_Z \cos \theta_W \sin \beta & -\mu &
0}\right).
\ee

Thus the cascade decay involves a host of new parameters
($M_1,M_2,\mu$ and $\tan\beta$) along with the parent squark or
gluino mass.  Consequently one looks for theoretical constraints
relating these parameters to one another.
\bigskip

\noindent {\bf The GUT and SUGRA Constraints:}  The most important
theoretical constraint comes from GUT, which implies the famous
unification of $SU(3) \times SU(2) \times U(1)$ gauge couplings
(Fig. 1).  It also implies the unification of the corresponding
gaugino masses at the GUT scale, since they evolve exactly like the
gauge couplings, i.e.
\be
M_i (\mu) = m_{1/2} \alpha_i (\mu)/\alpha_i (M_G).
\ee
Thus one sees from the Fig. 1 [5], that at the low-energy scale, $\mu
\sim M_W$,
\beq
M_2 &=& m_{1/2} \alpha_2/\alpha_2 (M_G) \simeq m_{1/2}, \nonumber
\\[2mm] M_1 &=& M_2 (\alpha_1/\alpha_2) \simeq M_2/2, \nonumber
\\[2mm] M_{\tilde g} &\equiv& M_3 = M_2(\alpha_3/\alpha_2) \simeq
3M_2.
\eeq

Unlike the gaugino mass unification, the unification of scalar masses
at $M_G$ does not follow from any GUT symmetry but from the minimal
SUGRA model. As per this model, SUSY is broken in the hidden sector
and its effect communicated to the observable sector via gravitational
interaction.  Since this interaction is colour and flavour blind, it
leads to a common soft SUSY breaking mass for all the scalars
\be
{\cal L}_{\rm soft} = m^2_0 (\tilde\ell^2_i + \tilde q^2_i + H^2_k) +
\cdots .
\ee
In addition there is a supersymmetric contribution to the Higgs
masses, $\mu^2 H^2_k$, following from the superpotential $W = \mu H_1
H_2 + h_t Q H_2 U + h_b Q H_1 D + h_\tau L H_1 E$.  Thus the GUT scale
unification of scalar masses is admittedly a model dependent
assumption.  Nonetheless it makes a remarkable prediction when evolved
to low energy (Fig. 2) --- i.e. one of Higgs scalar masses,
$M^2_{H_2}$, is driven negative by the large top Yukawa coupling
contribution $h_t t\bar t H_2$ [6].  This is the famous radiative
mechanism of electroweak symmetry breaking.  Requiring this EWSB to
occur at the right mass scale determines the magnitude of $\mu$, i.e.
\be
|\mu| \sim (2-3) m_{1/2} \sim (2-3) M_2.
\ee
\bigskip

\noindent {\bf SUGRA Signals at LHC:}  It is clear from (8,10 and 12)
that the lighter chargino and neutralino 
states in the SUGRA model are dominated by gaugino components, 
\be
\chi^\pm_{1,2} \simeq \tilde W^\pm, \tilde H^\pm; \chi^0_{1,2,\cdots}
\simeq \tilde B, \tilde W^0, \cdots . 
\ee
Moreover one expects only a modest effect from the mixing between the
gaugino and Higgsino components and hence the relevant parameter,
$\tan\beta$. 

With the above systematics one can understand the essential features
of cascade decay.  For illustration I shall briefly discuss cascade
decay of gluino for two representative gluino mass regions of interest
to LHC.

\begin{enumerate}
\item[{i)}] $M_{\tilde g} \simeq 300 \GeV$: In this case the gluino
decays into the light quarks
\be
g \rightarrow \bar q q \left[\tilde B (.2), \tilde W^0 (.3). \tilde
W^\pm (.5)\right], \ q \neq t,
\ee
which have negligible Yukawa couplings.  Thus the decay branching
ratios are proportional to the squares of the respective gauge
couplings as indicated in parantheses.  Because of the smaller $U(1)$
gauge coupling relative to the $SU(2)$, the direct decay into the LSP
$(\tilde B)$ is small compared to cascade decay via the heavier
($\tilde W$ dominated) chargino and neutralino states.  The latter
decay into the LSP via real or virtual $W/Z$ emission,
\be
\tilde W_0 \rightarrow Z \tilde B \rightarrow \ell^+ \ell^- \tilde B
(.06), \ \tilde W^\pm \rightarrow W\tilde B \rightarrow \ell^\pm \nu
\tilde B (.2),
\ee
whose leptonic branching ratios are indicated in parantheses.  From
(14) and (15) one can easily calculate the branching ratios of
dilepton and trilepton states resulting from the decay of a gluino
pair (5).  In particular the dilepton final state via charginos has a
branching ratio of 
$1\%$.  Then the Majorana nature of $\tilde g$ implies a distinctive
like sign dilepton (LSD) signal with a BR of $\sim 1/2\%$.

\item[{ii)}] $\tilde M_{\tilde g} \gsim 500 \GeV$: In this case the
large top Yukawa coupling implies a significant decay rate via
\be
\tilde g \rightarrow t \bar b \tilde H^-,
\ee
where both $t$ and $\tilde H^-$ can contribute to the leptonic final
state via
\be
t \rightarrow bW^+ \rightarrow b \ell^+ \nu (.2), \ \tilde H^-
\rightarrow W^- \tilde B \rightarrow \ell^- \nu \tilde B (.2).
\ee
Consequently the BR of the LSD signal from the decay of the gluino
pair is expected to go up to $2-3\%$.
\end{enumerate}

Fig. 3 shows the expected LSD signal from gluino pair production at
LHC for $M_{\tilde g} = 300$ and $800 \GeV$ along with the background
[7].  The latter comes from $\bar tt$ via cascade decay (long dashed)
or charge misidentification (dots).  Note that the signal is
accompanied by a much larger $p\!\!\!/_T$ compared to the background
because of the LSPs.  This can be used to effectively suppress the
background while retaining about $1/2$ of the signal.  Consequently
one can search for a gluino upto at least $800 \GeV$ at the low
luminosity $(10 fb^{-1})$ run of LHC, going upto $1200 \GeV$ at the
high luminosity $(100 fb^{-1})$.

Fig. 4 shows the size of the canonical $p\!\!\!/_T +$ jets signal
against gluino mass for two cases -- $M_{\tilde g} \ll M_{\tilde q}$
(triangles) and $M_{\tilde g} \simeq M_{\tilde q}$ (squares) [8].
The background line shown also corresponds to $5\sqrt{B}$ for the
LHC luminosity of $10 fb^{-1}$.  Thus one expects a $5\sigma$
discovery limit of at least upto $M_{\tilde g} = 1300 \GeV$ from this
signal.  Finally Fig. 5, shows the CMS
simulation [9] for the $5\sigma$ discovery limits from the various
leptonic channels in the plane of $m_0 - m_{1/2}$, the common scalar
and gaugino masses at the GUT scale.  The corresponding squark and
gluino mass contours are also shown.  As we see from this figure, it
will be possible to extend the squark and gluino searches at LHC well
into the TeV region.   
\bigskip

\noindent {\bf Relaxing the SUGRA Constraint:} As mentioned earlier,
the GUT scale unification of scalar masses (11) is highly model
dependent.  Even in SUGRA model it can be broken by nonminimal
contributions to ${\cal L}_{\rm soft}$.  Alternatively, starting from the
minimal SUGRA constraint (11) at the Plank scale one can get large
splitting between the soft masses of $\tilde q$, $\tilde \ell$ and the
Higgs scalars at the GUT scale, since they follow different evolution
equations [10].  Therefore it is important to probe for the SUSY
signal by relaxing the SUGRA constraints (11,12).  Its main effect on
the gluino signature comes from floating the $\mu$ parameter and in
particular allowing the light Higgsino region, 
\be
|\mu| \lsim M_{1,2}.
\ee
Here one expects a near degeneracy between the $\chi^0_1$ and $\chi^0_2$
masses $(\simeq |\mu|)$ in contrast to a factor of 2 difference
between them in the SUGRA case (12,13).  Thus one expects very
different event kinematics in the two regions.  Moreover the most
important gluino decays in the light Higgsino region are
\be
\tilde g \rightarrow t \bar b \chi^\pm_1, \ t\bar t \chi^0_1
\ee
which result in a significantly larger LSD signal compared to the
SUGRA case [7].

It should be noted that the SUSY search at LEP has been carried out
over the full $\mu - M_2$ plane.  In contrast the investigations for
hadron colliders have largely been restricted to the region of SUGRA
constraint (12), i.e. the light gaugino region (13).  This was partly
a matter of expediency for Tevatron, since in the complimentary region
of light Higgsino (18) the LEP search via $Z \rightarrow \chi^0_1
\chi^0_1$ had already pre-emptied the range $M_2 \rightarrow 100 \GeV$
(i.e. $M_{\tilde g} \rightarrow 300 \GeV$).  Nontheless it is
desirable to cover this range via the direct gluino search at
Tevatron.  There is of course no LEP constraint for the $M_{\tilde g}$
range of interest to LHC.  Therefore the SUSY simulations for LHC
should be extended to cover at least some representative values of
$\mu$ in the $|\mu| \lsim M_{1,2}$ region [7].
\bigskip\bigskip

\noindent {\bf References}:
\medskip

\begin{enumerate}
\item[{1.}] H.P. Nilles, Phys. Rep. 110, 1 (1984); H.E. Haber and
G.L. Kane, Phys. Rep. 117, 75 (1985).
\item[{2.}] G.L. Kane and J.P. Leville, Phys. Lett. B112, 227 (1982);
P.R. Harrison and C.H. Llewellyn-Smith, Nucl. Phys. B213, 223 (1983)
[Err. Nucl. Phys. B223, 542 (1983)]; E. Reya and D.P. Roy,
Phys. Rev. D32, 645 (1985).
\item[{3.}] W. Beenakker, R. Hopker, M. Spira and P. Zerwas,
Nucl. Phys. B492, 51 (1997); M. Kramer, T. Plehn, M. Spira and
P. Zerwas, Phys. Rev. Lett. 79, 341 (1997).
\item[{4.}] CDF and ${\rm DO\!\!\!\!/}$ collaborations: R. Culbertson,
Fermilab-Conf-97-277-E, Proc. SUSY97.
\item[{5.}] V. Barger, M.S. Berger and P. Ohmann, Phys. Rev. D47, 1093
(1993).
\item[{6.}] G.G. Ross and R.G. Roberts, Nucl. Phys. B377, 571 (1992);
see also M. Drees and M.M. Nojiri, Nucl. Phys. B369, 54 (1992).
\item[{7.}] M. Guchait and D.P. Roy, Phys. Rev. D52, 133 (1995).
\item[{8.}] H. Baer, C. Chen, F. Paige and X. Tata, Phys. Rev. D52,
2746 (1995). 
\item[{9.}] CMS Collaboration: D. Dengeri (Private communication).
\item[{10.}] N. Polansky and A. Pomarol, Phys. Rev. Lett. 73, 2292
(1994); Phys. Rev. D51, 6532 (1995); Y. Kawamura, H. Murayama and
M. Yamaguchi, Phys. Lett. B324, 54 (1994); Phys. Rev. D51, 1337 (1995).
\end{enumerate}

\newpage

\begin{figure}[htb]
\begin{center}
\leavevmode
\hbox{%
\epsfxsize=3.5in
\epsffile{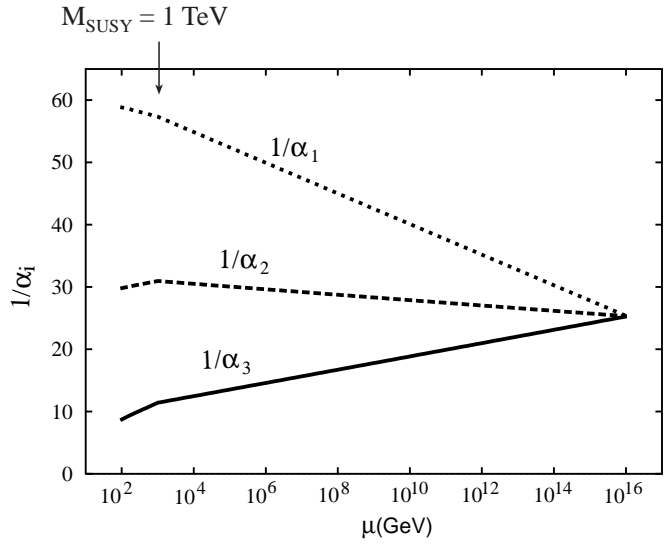}}
\caption{Unification of the $SU(3) \times SU(2) \times U(1)$ gauge
couplings at $M_G$ in the SUSY-GUT model [5].} 
\label{fig:susy1}
\end{center}
\end{figure}

\begin{figure}[htb]
\begin{center}
\leavevmode
\hbox{%
\epsfxsize=3.5in
\epsffile{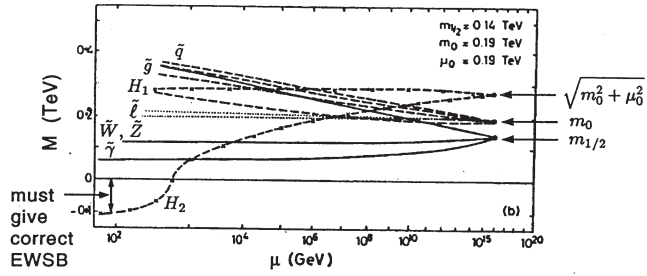}}
\caption{Evolution of the scalar and the gaugino masses in the SUGRA
model [6].  The negative $M_{H_2}$ line actually represents
$-|M_{H_2}|$ since the $M^2_{H_2}$ becomes negative in this region.
The supersymmetric Higgs mass parameter is denoted by $\mu_0$.}
\label{fig:susy2}
\end{center}
\end{figure}

\newpage

\begin{figure}[htb]
\begin{center}
\leavevmode
\hbox{%
\epsfxsize=3in 
\epsffile{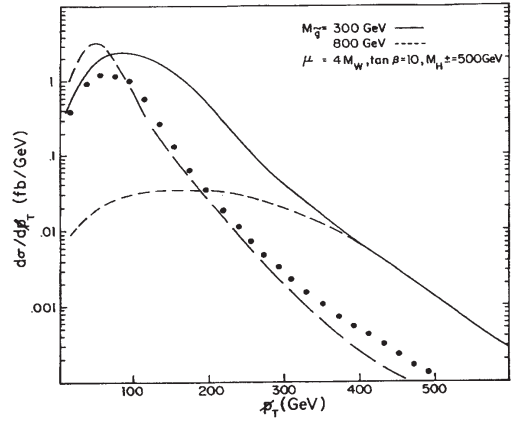}}
\caption{The expected size of the LSD signals for 300 and 800 GeV
gluino production at LHC are shown against the accompanying
missing-$p_T$.  The real and fake LSD backgrounds from $\bar tt$
production are shown by long dashed and dotted lines respectively [7].} 
\label{fig:susy3}
\end{center}
\end{figure}

\includegraphics{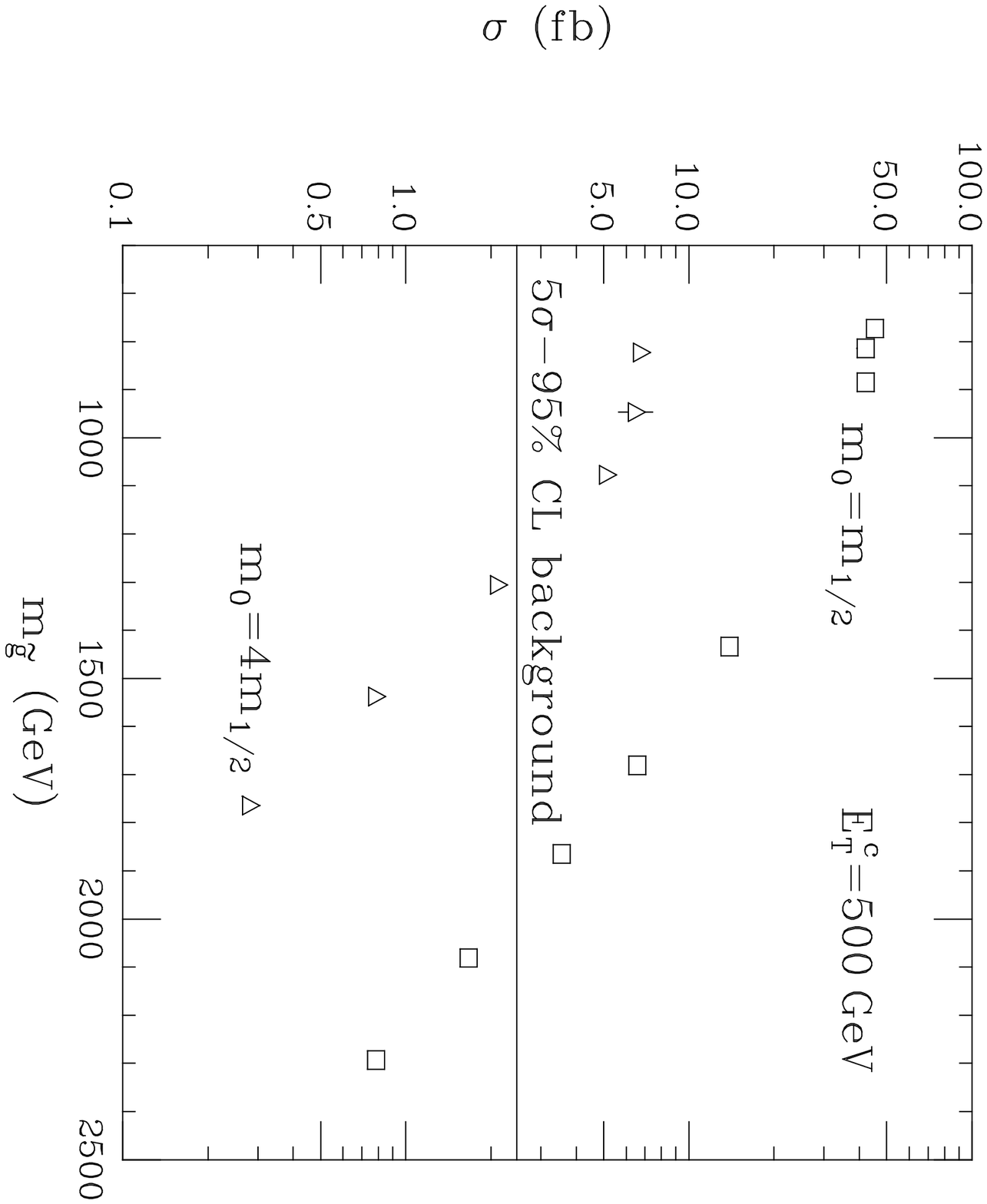}
\label{fig:susy4.eps}
\begin{figure}
\caption{The expected gluino signals at LHC from jets + missing-$E_T$
channel are shown for $M_{\tilde g} \simeq M_{\tilde q}$ (squares) and
$M_{\tilde g} \ll M_{\tilde q}$ (triangles).  The 95\% CL background
shown also corresponds to $5\sqrt{B}$ for the LHC luminosity of
$10fb^{-1}$ [8].}
\end{figure}

\newpage

\begin{figure}[htb]
\begin{center}
\leavevmode
\hbox{%
\epsfxsize=4in
\epsffile{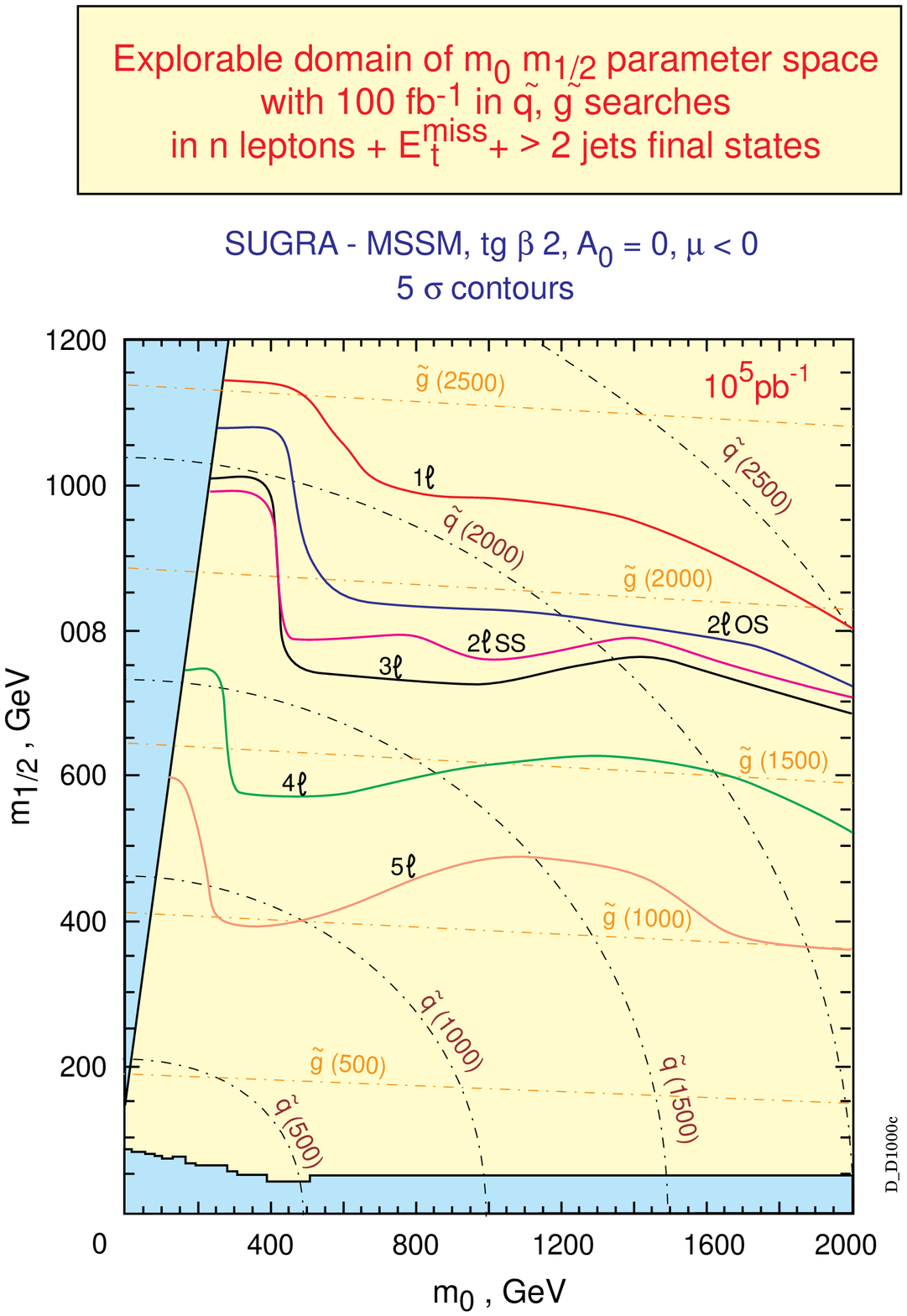}}
\caption{The SUSY discovery limits of various leptonic channels at
LHC, where $2l OS$ and $2l SS$ denote opposite sign and same sign
dileptons [9].}
\label{fig:susy5}
\end{center}
\end{figure}

\end{document}